\documentclass[pdflatex,sn-mathphys-num]{sn-jnl}

\usepackage{graphicx}%
\usepackage{multirow}%
\usepackage{amsmath,amssymb,amsfonts}%
\usepackage{amsthm}%
\usepackage{mathrsfs}%
\usepackage[title]{appendix}%
\usepackage{xcolor}%
\usepackage{textcomp}%
\usepackage{manyfoot}%
\usepackage{booktabs}%
\usepackage{algorithm}%
\usepackage{algorithmicx}%
\usepackage{algpseudocode}%
\usepackage{listings}%

\newcommand{\ket}[1]{|#1\rangle}
\newcommand{\bra}[1]{\langle#1|}

\newcommand{\ketbra}[2]{|#1\rangle\langle#2|}
\newcommand{\ave}[1]{\langle #1 \rangle}

\begin{document}

\title{Parrondo's paradox for three discrete-time quantum walks with distinct shift operators}

\author[1]{\fnm{Zbigniew} \sur{Walczak}}\email{zbigniew.walczak@uni.lodz.pl}

\author*[1]{\fnm{Jaros{\l}aw H.} \sur{Bauer}}\email{jaroslaw.bauer@uni.lodz.pl}

\affil[1]{\orgdiv{Department of Theoretical Physics, Faculty of Physics and Applied Informatics}, \orgname{University of Lodz}, \orgaddress{\street{Pomorska 149/153}, \city{90-236 Lodz}, \country{Poland}}}

\abstract{Parrondo's paradox describes a surprising phenomenon in which certain combinations of biased quantum walks result in an unexpected reversal of their bias direction. To date, research on this effect has predominantly focused on one-dimensional discrete-time quantum walks that employ fixed sequences of two or three quantum coins, paired with a single shift operator. In this work, we first introduce a wide class of unitary shift operators which allow for three movement options (left, right or stay in place) of a walker in one time step, and we then demonstrate that Parrondo's paradox can also arise with a single coin and three distinct shift operators. Furthermore, we investigate how this paradox influences the time evolution of quantum entanglement between the walker and the coin in such quantum walks.}


\maketitle

\section{Introduction}
While classical random walks have a rich and established history, serving as a foundational mathematical tool across diverse scientific disciplines, their quantum counterparts represent a more recent yet rapidly advancing development \cite{Aharonov93,Farhi98}.
Classical walks are inherently governed by stochastic processes and diffusive dynamics, which have proven invaluable in algorithmic design for complex problem-solving, graph traversal, and statistical physics. 
In contrast, quantum walks exploit non-classical phenomena - most notably unitary time evolution and quantum interference - which dramatically reshape the probability distribution of the walker. This fundamental departure from classical dynamics enables ballistically spreading wavepackets, thereby offering significant computational speedups in various quantum algorithms (e.g. spatial search and element distinctness) and paving the way for sophisticated quantum simulations of complex physical systems \cite{Aharonov93,Farhi98,Ambainis01,Konno02}. Driven by this theoretical advantages, quantum walks have been realized experimentally across a broad range of physical platforms, including trapped atoms and ions, integrated photonic waveguides and free-space optical systems, superconducting qubits, and nuclear magnetic resonance (NMR) setups, fueling intense multidisciplinary research over the past few decades \cite{Kempe03,Venegas-Andraca12,Manouchehri14,Portugal18,Neves18,Kadian21}.

A particularly intriguing concept originating from game theory is the counterintuitive phenomenon known as Parrondo's paradox \cite{Parrondo96,Harmer99b}, wherein a suitable combination of individualy losing strategies unexpectedly yields a winning outcome. In the classical realm, this effect finds a rigorous mathematical explanation through time-dependent or state-dependent Markov chains \cite{Parrondo00,Parrondo04,Dinis08}. Beyond its original game-theoretic formulation \cite{Parrondo00,Toral01,Dinis03,Kay03,Parrondo04,Dinis08,Khan10,Xie11,Ye13,Cheung16,Cheong17,Koh19,Koh20,Ye21,Lai22b,
Miszczak22}, Parrondo's paradox (or Parrondo's effect) has profoundly influenced a wide spectrum of fields, including nonlinear dynamics, biophysics, evolutionary biology, and statistical mechanics. Recently, at the microscopic quantum level, the flashing-ratchet effect in directional transport in organic devices has been demonstrated \cite{Li25,Li25b,Li26}. Comprehensive reviews on Parrondo's effect can be found in Refs.~\cite{Harmer02,Abbott10,Cheong19,Lai20d,Lai20c}.

In quantum information science, particularly within the framework of discrete-time coined quantum walks (DTQWs) on a one-dimensional lattice, Parrondo's effect has attracted considerable attention and has been extensively investigated in 
Refs.~\cite{Khan10,Meyer02,Flitney04,Kosik07,Bulger08,Chandrashekar11,Flitney12,Li13,Pawela13,Machida18,Rajendran18,Rajendran18b,
Jan20,Jan20b,Lai20,Lai20b,Pires20,Lai21,Panda21,Walczak21,Naves22,Panda22,Trautmann22,Walczak22,Fang23,Jan23,Mielke23,
Naves23,Walczak23,Kadiri24,Mittal24,Cavazzoni24,Mishra24,Razzoli24,Ximenes24,Walczak25}. 
Standard DTQWs are bipartite quantum systems consisting of a coin and a walker moving on the lattice \cite{Aharonov93,Meyer96,Portugal18}. The single-step time evolution is traditionally governed by a global unitary operator composed of a uniform coin operator followed by a conditional shift operator. 
However, non-unitary shift operators are sometimes employed instead \cite{Mendonca20,Gong22,Bauer24,Huo24}.
While previous studies of Parrondo's effect in DTQWs have focused almost exclusively on systems employing multiple alternating coin operators together with a single shift operator, the present work departs from this paradigm by adopting a novel framework, introduced in Ref.~\cite{Walczak24}, in which these roles are reversed. Specifically, we investigate DTQWs driven by a single fixed coin operator combined with multiple sequence-dependent shift operators.

In this work, we introduce a broad class of unitary shift operators that allow three possible movements at each time step: the walker may move one unit to the left, move one unit to the right, or remain stationary. We also formulate and prove sufficient conditions for the existence of such generalized unitary shift operators. As a consequence, we explicitly demonstrate that Parrondo's effect can arise not only in DTQWs with a single coin operator and two distinct shift operators \cite{Walczak24}, but also in a more general setting involving three distinct shift operators.
Moreover, we investigate how Parrondo's effect influences the time evolution of the walker-coin quantum entanglement and walker spreading.

\section{DTQWs}
A DTQW (with sequences of two-state quantum coins) is defined on the Hilbert space ${\cal H} = {\cal H}_{w} \otimes {\cal H}_{c}$, where ${\cal H}_{w}$ is the walker Hilbert space spanned by the orthonormal states $\{ \ket{x} \colon x \in \mathbb{Z} \}$ representing the walker's position on the lattice, and ${\cal H}_{c}$ is the coin Hilbert space spanned by the orthonormal states 
$\{ \ket{0}, \ket{1} \}$. A separable state of the form   
\begin{align}
\label{IS}
\ket{\psi(0)} = \ket{0} \otimes \ket{0}
\end{align}
is the standard initial state of the walker-coin system. At time step $t$, the state of the walker-coin system can be expressed as \cite{Portugal18}
\begin{align}
\label{psit}
\ket{\psi(t)} = \sum_{x} \ket{x} \otimes [a(x,t) \ket{0} + b(x,t) \ket{1}],
\end{align}
where $a(x,t)$ and $b(x,t)$ are the probability amplitudes satisfying the standard normalization condition
\begin{align}
\label{NC}
\sum_{x} [|a(x,t)|^{2} + |b(x,t)|^{2}] = 1.
\end{align} 
The one-step time evolution of the walker-coin system
\begin{align}
\label{psit1}
\ket{\psi(t+1)} = U \ket{\psi(t)}
\end{align}
is, in general, governed by the unitary operator
\begin{align}
\label{U}
U = S ({I}_{w} \otimes C),
\end{align}
where $C$ is the (unitary) coin operator acting on ${\cal H}_{c}$, ${I}_{w}$ is the identity operator acting on ${\cal H}_{w}$, and $S$ is the (unitary) conditional shift operator acting on ${\cal H}$. In this work, the Hadamard coin operator
\begin{align}
\label{Had}
C_{H} = \frac{1}{\sqrt{2}}
\begin{pmatrix}
1 & ~~1 \\
\rule{0mm}{5mm}
1 & -1
\end{pmatrix}
\end{align}
is used in place of $C$.
Let us introduce the following notation
\begin{align}
\label{SPM}
{S}_{-} = \sum_{x} \ketbra{x - 1}{x}, \qquad {S}_{+} = \sum_{x} \ketbra{x + 1}{x}.
\end{align}
Let us note that identity operators $I_{w}$ and $I_{c}$ are given by
\begin{align}
\label{IwIc}
{I}_{w} = \sum_{x} \ketbra{x}{x}, \qquad {I}_{c} = \ketbra{0}{0} + \ketbra{1}{1},
\end{align}
respectively.
The following standard unitary shift operator was employed in our previous work on Parrondo's effect \cite{Walczak21,Walczak22,Walczak23} 
\begin{align}
\label{S}
S = {S}_{-} \otimes \ketbra{0}{0} + {S}_{+} \otimes \ketbra{1}{1}.
\end{align}
This operator allows the walker to move one unit left or right (in the $x$ direction) after one time step. 
In the present work we will also use two other unitary shift operators, namely
\begin{align}
\label{T}
T(\alpha) & = \sin(\alpha) ~{S}_{+} \otimes \ketbra{1}{0} - 
              \sin(\alpha) ~{S}_{-} \otimes \ketbra{0}{1} + \cos(\alpha) ~{I}_{w} \otimes I_{c}, \\
\label{V}
V(\beta) & = \cos(\beta) ~{S}_{-} \otimes \ketbra{0}{0} + 
             \cos(\beta) ~{S}_{+} \otimes \ketbra{1}{1} + \sin(\beta) ~{I}_{w} \otimes (\ketbra{0}{1}-\ketbra{1}{0}),
\end{align}
which depend on two real parameters $\alpha$ and $\beta$, respectively. Note that $S = V(0)$. The operators from Eqs. (\ref{T}) and (\ref{V}) allow the walker to move one unit left or right but the walker can also remain in its starting position after one time step. As far as we know, unitary shift operators with this property have only recently come into use in Ref.~\cite{Passos22} (and references therein). Both $T(\alpha)$, $V(\beta)$ and the shift operators from Refs.~\cite{Passos22,Walczak24} belong to a wider class of shift operators which have the above-mentioned properties. This class of shift operators will be described in the next section.

\section{General form of the shift operator}
In most works on DTQWs, shift operators do not change the state of the coin after one time step in the walker-coin system evolution. This means that only terms with $\ketbra{0}{0}$ or $\ketbra{1}{1}$ on the right-hand side of the tensor products may be present in the shift operator $S$ (cf. Eq. (\ref{S})). In this section, we will show that terms with $\ketbra{0}{1}$ or $\ketbra{1}{0}$ in the shift operator are also necessary (cf. Eqs. (\ref{T}) and (\ref{V})) if two conditions are to be met simultaneously. Namely, unitarity of the shift operator (which ensures that $U$ is unitary) and the possibility that the walker can stay put or move one unit left or right after one time step, are these conditions. Let us note that DTQWs with shift operators of this type differ from standard DTQWs (with $S$ from Eq. (\ref{S}) or similar) because in the former the states $\{ \ket{x} \colon x \in \mathbb{Z} \}$ with both odd and even numbers are occupied at each time step. 
The most general form of the shift operator which does not change the state of the coin and allows for the possibility that the walker can stay put or move one unit left or right after one time step may be written as follows
\begin{align}
\label{S1}
\tilde{S} & = a{S}_{-} \otimes \ketbra{0}{0} + b{S}_{+} \otimes \ketbra{0}{0} + c{I}_{w} \otimes \ketbra{0}{0} \nonumber\\
  & + a'{S}_{-} \otimes \ketbra{1}{1} + b'{S}_{+} \otimes \ketbra{1}{1} + c'{I}_{w} \otimes \ketbra{1}{1},       
\end{align}
where $a, b, c, a', b', c' \in \mathbb{C}$. These 6 complex numbers should be chosen in such a way that
\begin{align}
\label{S2}
\tilde{S} \tilde{S}^{\dagger} = \tilde{S}^{\dagger} \tilde{S} = {I}_{w} \otimes {I}_{c}.
\end{align}
($\tilde{S}^{\dagger}$ denotes the Hermitian conjugate of $\tilde{S}$.) Based on Eqs. (\ref{SPM}) and (\ref{IwIc}), one can derive the following relationships
\begin{align}
\label{S3}
{S}_{-}{S}_{+} = ~ {S}_{+}{S}_{-} = {I}_{w}, \qquad {S}_{-}^{\dagger} = ~ {S}_{+}, \qquad {S}_{+}^{\dagger} = ~ {S}_{-}.
\end{align}
Hence, based on Eqs. (\ref{S3}), one easily computes that
\begin{align}
\label{S4}
\tilde{S}^{\dagger} & = a^{\ast}{S}_{+} \otimes \ketbra{0}{0} + b^{\ast}{S}_{-} \otimes \ketbra{0}{0} + c^{\ast}{I}_{w} \otimes \ketbra{0}{0} \nonumber\\
  & + a'^{\ast}{S}_{+} \otimes \ketbra{1}{1} + b'^{\ast}{S}_{-} \otimes \ketbra{1}{1} + c'^{\ast}{I}_{w} \otimes \ketbra{1}{1}.       
\end{align}
(The asterisk denotes the complex conjugate.) Since ${S}_{-}{S}_{-} = {S}_{-}^{2}$, ${S}_{+}{S}_{+} = {S}_{+}^{2}$ and since the coin states $\{ \ket{0}, \ket{1} \}$ are orthonormal, one obtains
\begin{align}
\label{S5}
\tilde{S} \tilde{S}^{\dagger} & = [(|a|^{2}+|b|^{2}+|c|^{2}){I}_{w} + ab^{\ast}{S}_{-}^{2} + (ac^{\ast} 
+ b^{\ast}c){S}_{-} + (bc^{\ast} + a^{\ast}c){S}_{+} \nonumber\\
& + a^{\ast}b{S}_{+}^{2}] \otimes \ketbra{0}{0} \nonumber\\
& + [(|a'|^{2}+|b'|^{2}+|c'|^{2}){I}_{w} + a'b'^{\ast}{S}_{-}^{2} + (a'c'^{\ast} + b'^{\ast}c'){S}_{-}
+ (b'c'^{\ast} + a'^{\ast}c'){S}_{+} \nonumber\\
& + a'^{\ast}b'{S}_{+}^{2}] \otimes \ketbra{1}{1}.
\end{align}
To fulfill the relationship $\tilde{S} \tilde{S}^{\dagger} = {I}_{w} \otimes {I}_{c}$ in Eq. (\ref{S5}), the following conditions must be met: the coefficient multiplying ${I}_{w}$ must be equal to unity (in both terms: the one with $\ketbra{0}{0}$ and the one with $\ketbra{1}{1}$) and all the rest coefficients multiplying other operators must vanish (in both terms). Since there are 10 coefficients in Eq. (\ref{S5}), one obtains 10 equations for the 6 complex numbers $a, b, c, a', b', c'$. Among these equations, only 6 are independent and we show them below
\begin{align}
\label{S6}
& |a|^{2}+|b|^{2}+|c|^{2} = 1, \qquad ab^{\ast} = 0, \qquad ac^{\ast} + b^{\ast}c = 0, \\
\label{S6a}
& |a'|^{2}+|b'|^{2}+|c'|^{2} = 1, \qquad a'b'^{\ast} = 0, \qquad a'c'^{\ast} + b'^{\ast}c' = 0.
\end{align}
The condition $\tilde{S}^{\dagger} \tilde{S} = {I}_{w} \otimes {I}_{c}$ leads to the same system of equations. From Eqs. (\ref{S6}), it follows that exactly one number is nonzero among the numbers $a, b, c$. Likewise, from Eqs. (\ref{S6a}), it follows that exactly one number is nonzero among the numbers $a', b', c'$. As a result, there are exactly two nonzero terms in Eq. (\ref{S1}). Therefore, there is no unitary shift operator in the form of Eq. (\ref{S1}) that could complete all three movement options (mentioned above and represented by the operators ${S}_{-}$, ${S}_{+}$ and ${I}_{w}$, respectively) of the quantum walker after one time step.

Let us now consider a more general form of the shift operator by giving up the condition that only terms with $\ketbra{0}{0}$ or $\ketbra{1}{1}$ on the right-hand side of the tensor products may be present in the shift operator. This time, also terms with $\ketbra{0}{1}$ or $\ketbra{1}{0}$ can exist. Thus, the more general operator $\tilde{S}_{gen}$ is the following
\begin{align}
\label{S7}
\tilde{S}_{gen} & = a{S}_{-} \otimes \ketbra{0}{0} + b{S}_{+} \otimes \ketbra{0}{0} + c{I}_{w} \otimes \ketbra{0}{0} \nonumber\\
  & + a'{S}_{-} \otimes \ketbra{1}{1} + b'{S}_{+} \otimes \ketbra{1}{1} + c'{I}_{w} \otimes \ketbra{1}{1} \nonumber\\
  & + d{S}_{-} \otimes \ketbra{0}{1} + e{S}_{+} \otimes \ketbra{0}{1} + f{I}_{w} \otimes \ketbra{0}{1} \nonumber\\
  & + d'{S}_{-} \otimes \ketbra{1}{0} + e'{S}_{+} \otimes \ketbra{1}{0} + f'{I}_{w} \otimes \ketbra{1}{0}. 
\end{align}
where $a, b, c, a', b', c', d, e, f, d', e', f', \in \mathbb{C}$. These 12 complex numbers should be chosen in such a way that
\begin{align}
\label{S8}
\tilde{S}_{gen} \tilde{S}_{gen}^{\dagger} = \tilde{S}_{gen}^{\dagger} \tilde{S}_{gen} = {I}_{w} \otimes {I}_{c}.
\end{align}
Hence, based on Eqs. (\ref{S3}), one easily computes that
\begin{align}
\label{S9}
\tilde{S}_{gen}^{\dagger} & = a^{\ast}{S}_{+} \otimes \ketbra{0}{0} + b^{\ast}{S}_{-} \otimes \ketbra{0}{0} + c^{\ast}{I}_{w} \otimes \ketbra{0}{0} \nonumber\\
 & + a'^{\ast}{S}_{+} \otimes \ketbra{1}{1} + b'^{\ast}{S}_{-} \otimes \ketbra{1}{1} + c'^{\ast}{I}_{w} \otimes \ketbra{1}{1}       \nonumber\\
 & + d^{\ast}{S}_{+} \otimes \ketbra{1}{0} + e^{\ast}{S}_{-} \otimes \ketbra{1}{0} + f^{\ast}{I}_{w} \otimes \ketbra{1}{0} \nonumber\\
 & + d'^{\ast}{S}_{+} \otimes \ketbra{0}{1} + e'^{\ast}{S}_{-} \otimes \ketbra{0}{1} + f'^{\ast}{I}_{w} \otimes \ketbra{0}{1}.  
\end{align}
In the same way as before (that is, when calculating the right-hand side of Eq. (\ref{S5})), one obtains
\begin{align}
\label{S10}
\tilde{S}_{gen} \tilde{S}_{gen}^{\dagger} = {T}_{00} \otimes \ketbra{0}{0} + {T}_{01} \otimes \ketbra{0}{1} + {T}_{10} \otimes \ketbra{1}{0} + {T}_{11} \otimes \ketbra{1}{1},
\end{align}
where
\begin{align}
\label{S11}
{T}_{00} & = (|a|^{2}+|b|^{2}+|c|^{2}+|d|^{2}+|e|^{2}+|f|^{2}){I}_{w} + (cb^{\ast}+ac^{\ast}+df^{\ast}+fe^{\ast}){S}_{-} \nonumber\\
& + (ca^{\ast}+bc^{\ast}+ef^{\ast}+fd^{\ast}){S}_{+} + (ab^{\ast}+de^{\ast}){S}_{-}^{2} + (ba^{\ast}+ed^{\ast}){S}_{+}^{2},
\end{align}
\begin{align}
\label{S12}
{T}_{01} & = (ad'^{\ast}+be'^{\ast}+cf'^{\ast}+da'^{\ast}+eb'^{\ast}+fc'^{\ast}){I}_{w} + (af'^{\ast}+ce'^{\ast}+dc'^{\ast} + fb'^{\ast}){S}_{-} \nonumber\\
& + (bf'^{\ast}+cd'^{\ast}+ec'^{\ast}+fa'^{\ast}){S}_{+} + (ae'^{\ast}+db'^{\ast}){S}_{-}^{2} + (bd'^{\ast}+ea'^{\ast}){S}_{+}^{2},
\end{align}
\begin{align}
\label{S13}
{T}_{10} & = (a'd^{\ast}+b'e^{\ast}+c'f^{\ast}+d'a^{\ast}+e'b^{\ast}+f'c^{\ast}){I}_{w} + (a'f^{\ast}+c'e^{\ast}+d'c^{\ast} + f'b^{\ast}){S}_{-} \nonumber\\
& + (b'f^{\ast}+c'd^{\ast}+e'c^{\ast}+f'a^{\ast}){S}_{+} + (a'e^{\ast}+d'b^{\ast}){S}_{-}^{2} + (b'd^{\ast}+e'a^{\ast}){S}_{+}^{2},
\end{align}
\begin{align}
\label{S14}
{T}_{11} & = (|a'|^{2}+|b'|^{2}+|c'|^{2}+|d'|^{2}+|e'|^{2}+|f'|^{2}){I}_{w} + (c'b'^{\ast}+a'c'^{\ast}+d'f'^{\ast}+f'e'^{\ast}){S}_{-} \nonumber\\
& + (c'a'^{\ast}+b'c'^{\ast}+e'f'^{\ast}+f'd'^{\ast}){S}_{+} + (a'b'^{\ast}+d'e'^{\ast}){S}_{-}^{2} + (b'a'^{\ast}+e'd'^{\ast}){S}_{+}^{2}.
\end{align}
To fulfill the relationship $\tilde{S}_{gen} \tilde{S}_{gen}^{\dagger} = {I}_{w} \otimes {I}_{c}$ in Eq. (\ref{S10}), the following conditions must be met:
\begin{align}
\label{S15}
{T}_{00} = {I}_{w}, \qquad {T}_{01} = 0, \qquad {T}_{10} = 0, \qquad {T}_{11} = {I}_{w}.
\end{align}
As a result, the coefficients multiplying ${I}_{w}$ must be equal to unity in Eqs. (\ref{S11}) and (\ref{S14}), and must be equal to zero in Eqs. (\ref{S12}) and (\ref{S13}). Moreover, all the rest coefficients multiplying other operators (i.e., ${S}_{-}$, ${S}_{+}$, ${S}_{-}^{2}$, ${S}_{+}^{2}$, respectively) must vanish in Eqs. (\ref{S11})-(\ref{S14}). Since there are 20 such coefficients in Eqs. (\ref{S11})-(\ref{S14}), one obtains 20 equations for the 12 complex numbers $a, b, c, d, e, f, a', b', c', d', e', f'$. Among these equations, only 11 are independent, and we show them below
\begin{align}
\label{S16}
& |a|^{2} + |b|^{2} + |c|^{2} + |d|^{2} + |e|^{2} + |f|^{2} = 1, \\
\label{S17}
& ab^{\ast} + de^{\ast} = 0, \\
\label{S18}
& cb^{\ast} + ac^{\ast} + df^{\ast} + fe^{\ast} = 0, \\
\label{S19}
& |a'|^{2} + |b'|^{2} + |c'|^{2} + |d'|^{2} + |e'|^{2} + |f'|^{2} = 1, \\
\label{S20}
& a'b'^{\ast} + d'e'^{\ast} = 0, \\
\label{S21}
& c'b'^{\ast} + a'c'^{\ast} + d'f'^{\ast} + f'e'^{\ast} = 0, \\
\label{S22}
& ad'^{\ast} + be'^{\ast} + cf'^{\ast} + da'^{\ast} + eb'^{\ast} + fc'^{\ast} = 0, \\
\label{S23}
& af'^{\ast} + ce'^{\ast} + dc'^{\ast} + fb'^{\ast} = 0, \\
\label{S24}
& ae'^{\ast} + db'^{\ast} = 0, \\
\label{S25}
& bf'^{\ast} + cd'^{\ast} + ec'^{\ast} + fa'^{\ast} = 0, \\
\label{S26}
& bd'^{\ast} + ea'^{\ast} = 0.
\end{align}
The condition $\tilde{S}_{gen}^{\dagger} \tilde{S}_{gen} = {I}_{w} \otimes {I}_{c}$ leads to the same system of equations. There is a wide variety of shift operator ($\tilde{S}_{gen}$) choices available in accordance with Eqs. (\ref{S16})-(\ref{S26}). Two of these choices are shown in Eqs. (\ref{T}) and (\ref{V}). For the operator $T(\alpha)$, from Eq. (\ref{T}), the only nonzero coefficients in Eqs. (\ref{S16})-(\ref{S26}) are the following
\begin{align} 
\label{S27}
& c = \cos(\alpha), \qquad d = -\sin(\alpha), \qquad c' = \cos(\alpha), \qquad e' = \sin(\alpha).
\end{align}
For the operator $V(\beta)$, from Eq. (\ref{V}), the only nonzero coefficients in Eqs. (\ref{S16})-(\ref{S26}) are the following
\begin{align} 
\label{S28}
& a = \cos(\beta), \qquad f = \sin(\beta), \qquad b' = \cos(\beta), \qquad f' = -\sin(\beta).
\end{align}
One can easily check that Eqs. (\ref{S16})-(\ref{S26}) are satisfied in both cases. The shift operators $T(\alpha)$ and $V(\beta)$ have been constructed with maximum simplicity in mind, i.e. they should contain the fewest possible terms and implement the three motion options (mentioned above) for one time step each.

\section{Space-time evolution of the quantum walker}
\begin{figure}[t]
\centering
\begin{tabular}{cc}
\includegraphics[width=0.47\textwidth]{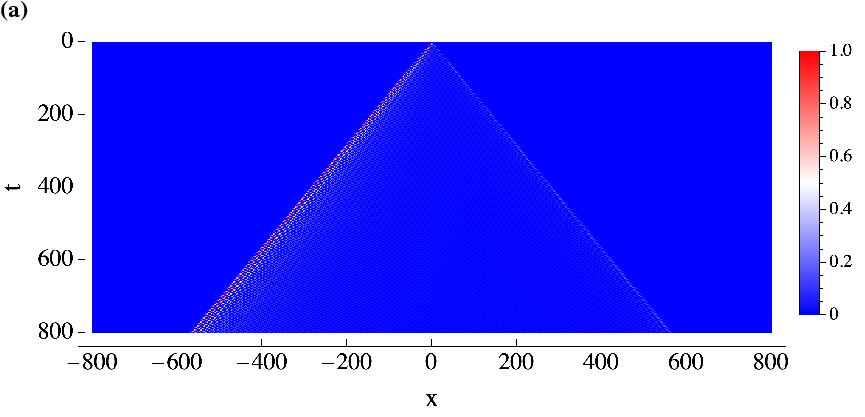} & \hspace{5mm} \includegraphics[width=0.47\textwidth]{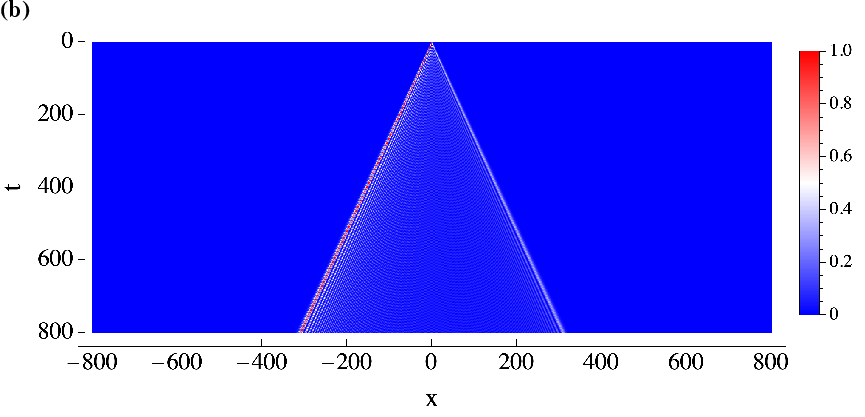} \vspace{5mm} \\
\includegraphics[width=0.47\textwidth]{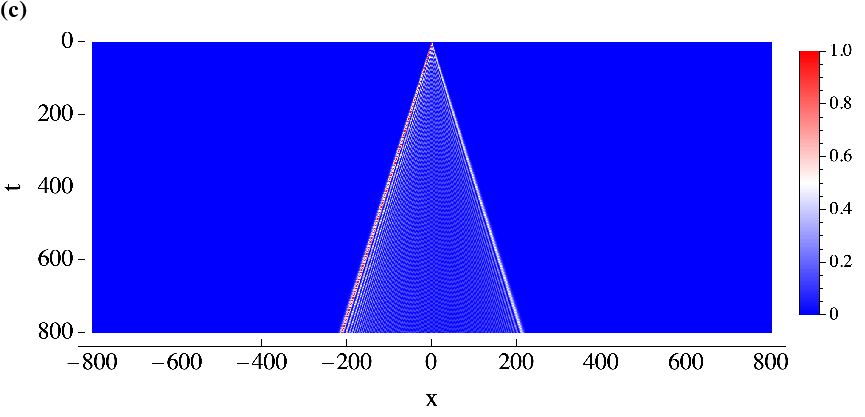} & \hspace{5mm} \includegraphics[width=0.47\textwidth]{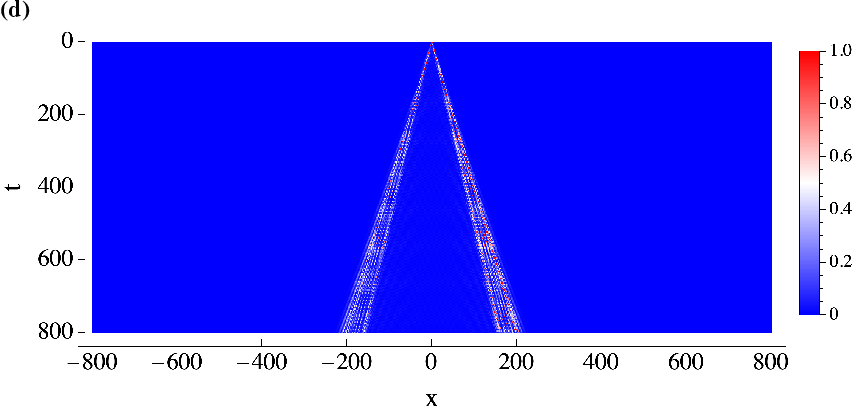} \vspace{5mm} \\
\includegraphics[width=0.47\textwidth]{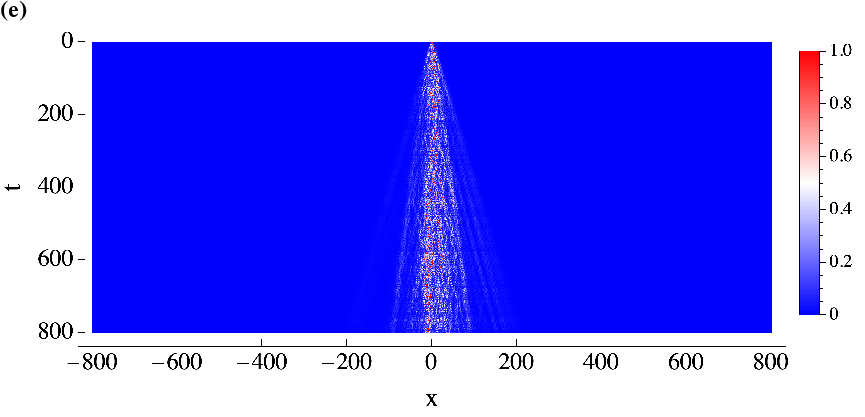}
\end{tabular}
\caption{The space-time evolution of $p(x, t)/\max_{x} p(x, t)$. The plots (a), (b), and (c) are for 
DTQWs with the shift operators $S$, $T$, and $V$, respectively, understood as DTQWs with the period-one sequence of the shift operators $S$, $T$, and $V$, respectively. The plots (d) and (e) are for DTQWs with deterministic periodic and aperiodic sequences of the shift operators $S$, $T$, and $V$, given by Eqs.~(\ref{DPSC}) and (\ref{DASC}), respectively.} 
\end{figure}

When the operator $T(\alpha)$ is applied, together with Eqs. (\ref{psit}) and (\ref{psit1})-(\ref{Had}), one can derive the following recurrence relations for the probability amplitudes $a(x,t)$ and $b(x,t)$
\begin{align}
a(x,t+1) & = \frac{1}{\sqrt{2}}[- \sin(\alpha)a(x+1,t) + \sin(\alpha)b(x+1,t) \nonumber\\
         &   + \cos(\alpha)a(x,t) + \cos(\alpha)b(x,t)], \\             
b(x,t+1) & = \frac{1}{\sqrt{2}}[\sin(\alpha)a(x-1,t) + \sin(\alpha)b(x-1,t) \nonumber\\ 
         &   + \cos(\alpha)a(x,t) - \cos(\alpha)b(x,t)].
\end{align}
On the other hand, when the operator $V(\beta)$ is applied with Eqs. (\ref{psit}) and (\ref{psit1})-(\ref{Had}), one obtains the following recurrence relations for the probability amplitudes $a(x,t)$ and $b(x,t)$
\begin{align}
a(x,t+1) & = \frac{1}{\sqrt{2}}[\cos(\beta)a(x+1,t) - \cos(\beta)b(x+1,t) \nonumber\\
         &   + i\sin(\beta)a(x,t) + i\sin(\beta)b(x,t)], \\             
b(x,t+1) & = \frac{1}{\sqrt{2}}[\cos(\beta)a(x-1,t) + \cos(\beta)b(x-1,t) \nonumber\\ 
         &   + i\sin(\beta)a(x,t) - i\sin(\beta)b(x,t)].
\end{align}

\begin{figure}[t]
\centering
\begin{tabular}{cc}
\includegraphics[width=0.47\textwidth]{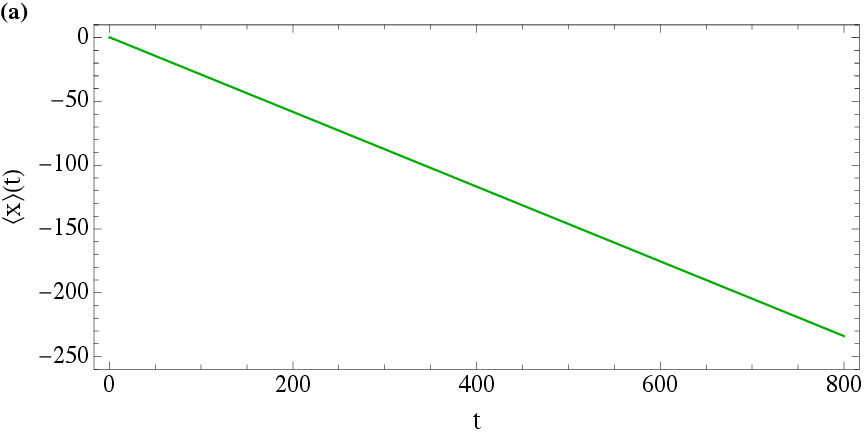} & \hspace{5mm} \includegraphics[width=0.47\textwidth]{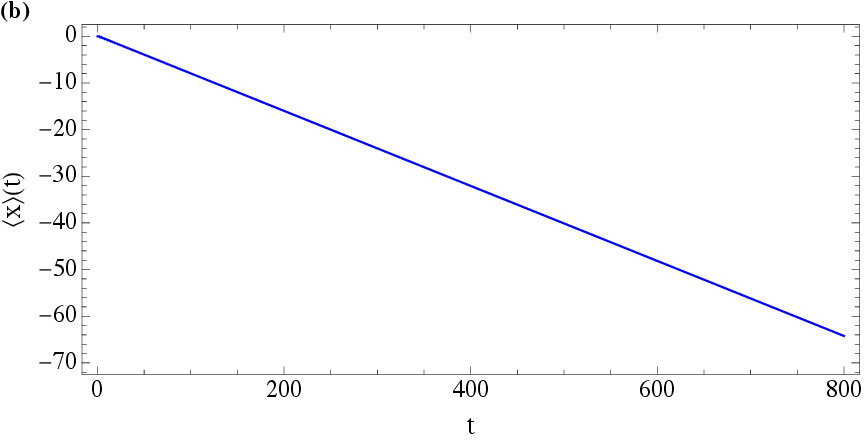} \vspace{5mm} \\
\includegraphics[width=0.47\textwidth]{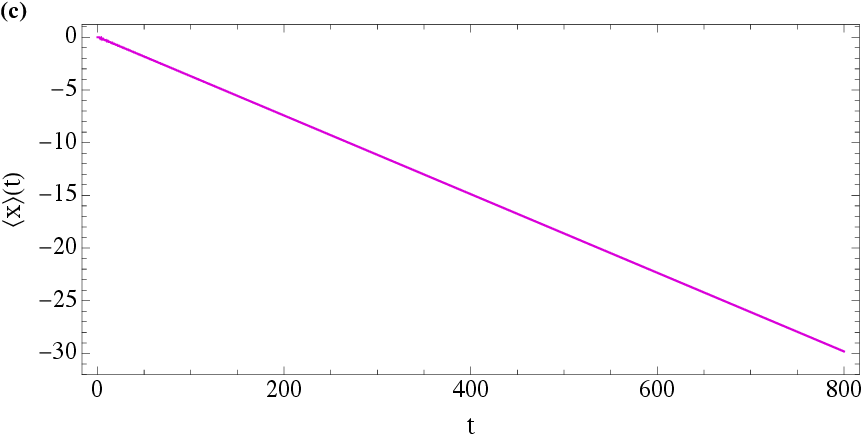} & \hspace{5mm} \includegraphics[width=0.47\textwidth]{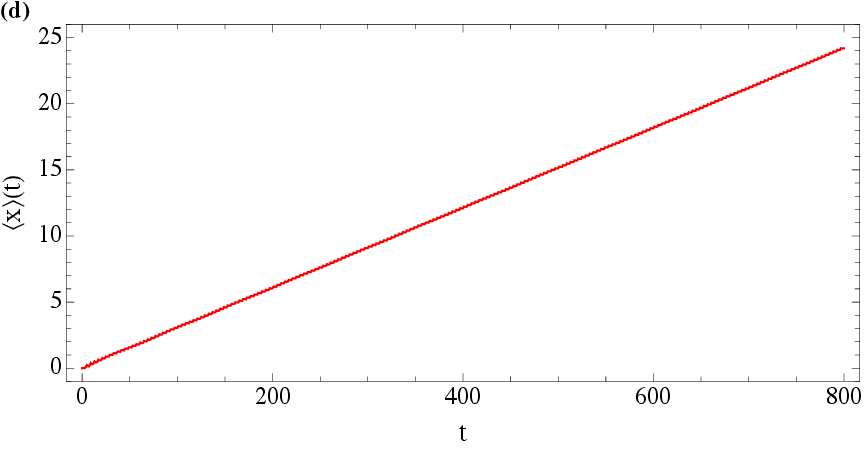} \vspace{5mm} \\
\includegraphics[width=0.47\textwidth]{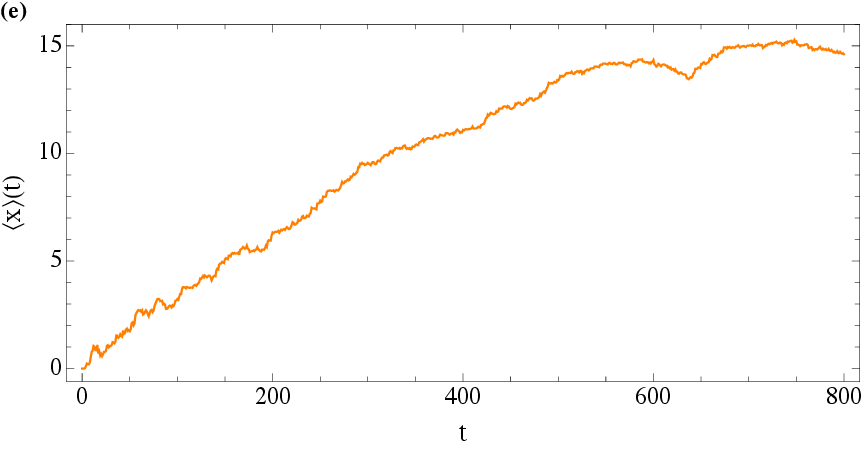}
\end{tabular}
\caption{The time evolution of the average position of the walker. The plots (a), (b) and (c) are for 
DTQWs with the shift operators $S$, $T$, and $V$, respectively, understood as DTQWs with the period-one sequence of the shift operators $S$, $T$, and $V$, respectively. The plots (d) and (e) are for DTQWs with deterministic periodic and aperiodic sequences of the shift operators $S$, $T$, and $V$, given by Eqs.~(\ref{DPSC}) and (\ref{DASC}), respectively.} 
\end{figure}

At time step $t$, the walker position probability distribution takes the form
\begin{align}
\label{PD}
p(x, t) = |a(x,t)|^{2} + |b(x,t)|^{2}.
\end{align}
The average position and the position variance of the walker are given by
\begin{align}
\label{AP} 
& \ave{x}(t) = \sum_{x} x p(x, t), \\
\label{VP}
& \sigma^{2}(t) = \sum_{x} x^{2} p(x, t) - \big( \sum_{x} x p(x, t) \big)^2,
\end{align}
respectively.
Since the state of the walker-coin system at time step $t$ is a pure quantum state, the walker-coin quantum entanglement at this time step can be quantified by entanglement entropy \cite{Abal06}
\begin{align}
\label{Sen}
{S}_{E}(t) = - \text{Tr}(\rho_{c}(t) \log_{2} \rho_{c}(t)),
\end{align}
where $\rho_{c}(t) = \text{Tr}_{w}(\ket{\psi(t)}\bra{\psi(t)})$ is the reduced density matrix describing the coin state at time step $t$ and $\text{Tr}_{w}$ denotes the partial trace over the walker subsystem.
It is worth noting here that the entanglement entropy ranges from $0$ for the walker-coin separable states to $1$ for the walker-coin maximally entangled states. 
At time step $t$, the coin state is given by \cite{Abal06}
\begin{align}
\rho_{c}(t) = 
\begin{pmatrix}
\rho_{11}(t) & \rho_{12}(t) \\
\rule{0mm}{5mm}
\rho_{12}^{\ast}(t) & 1 - \rho_{11}(t)
\end{pmatrix},
\end{align}
where $\rho_{11}(t) = \sum_{x} |a_{x}(t)|^{2}$ and $\rho_{12}(t) = \sum_{x} a_{x}(t) b_{x}^{\ast}(t)$.
The entanglement entropy (\ref{Sen}) can be expressed as \cite{Abal06}
\begin{align}
\label{Sen2}
{S}_{E}(t) = - \lambda_{+} \log_{2} \lambda_{+} - \lambda_{-} \log_{2} \lambda_{-},
\end{align}
where
\begin{align}
\lambda_{\pm} = \tfrac{1}{2} \pm \sqrt{\tfrac{1}{4} - \rho_{11}(t) [1 - \rho_{11}(t)] + |\rho_{12}(t)|^{2}}
\end{align}
are the eigenvalues of $\rho_{c}(t)$, and the standard convention that $0 \log_{2} 0 = 0$ is used.

\begin{figure}[t]
\centering
\begin{tabular}{cc}
\includegraphics[width=0.47\textwidth]{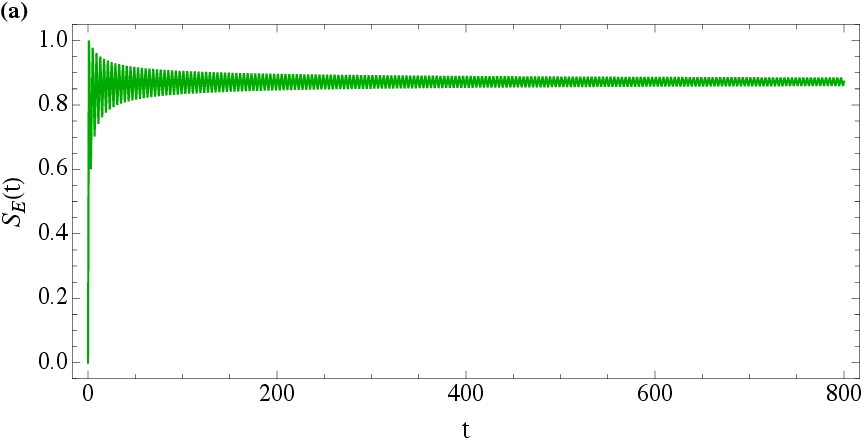} & \hspace{5mm} \includegraphics[width=0.47\textwidth]{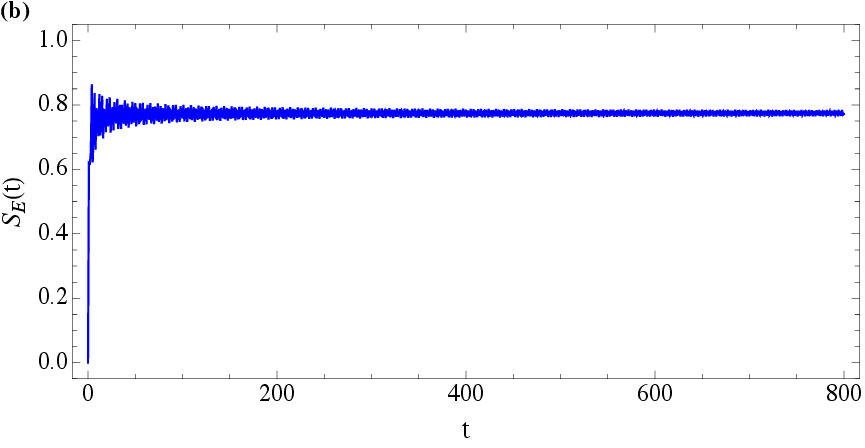} \vspace{5mm} \\
\includegraphics[width=0.47\textwidth]{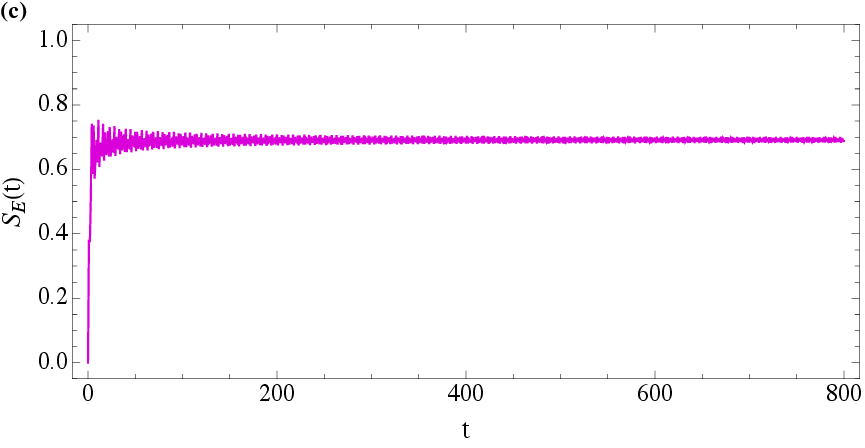} & \hspace{5mm} \includegraphics[width=0.47\textwidth]{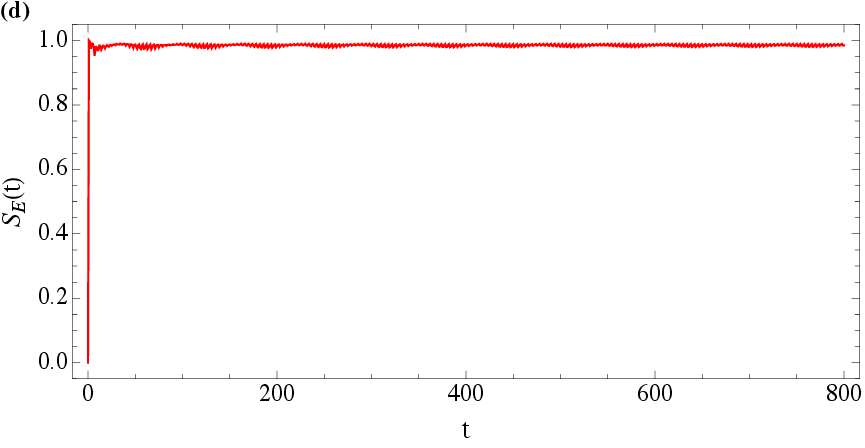} \vspace{5mm} \\
\includegraphics[width=0.47\textwidth]{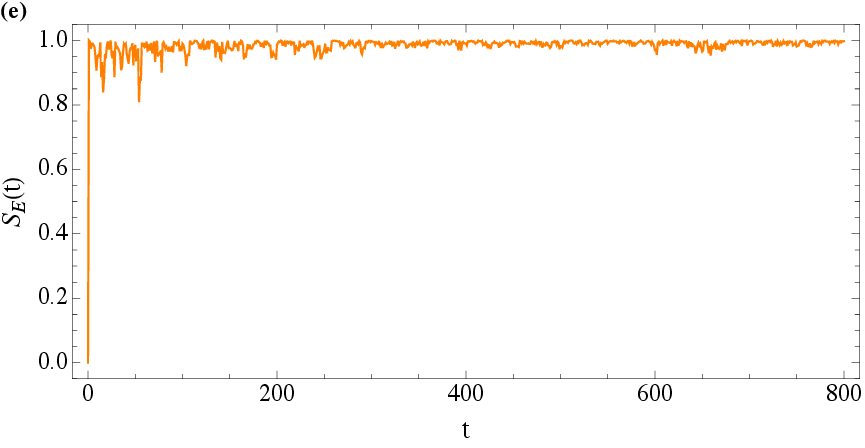}
\end{tabular}
\caption{The time evolution of the walker-coin quantum entanglement. The plots (a), (b), and (c) are for 
DTQWs with the shift operators $S$, $T$, and $V$, respectively, understood as DTQWs with the period-one sequence of the shift operators $S$, $T$, and $V$, respectively. The plots (d) and (e) are for DTQWs with deterministic periodic and aperiodic sequences of the shift operators $S$, $T$, and $V$, given by Eqs.~(\ref{DPSC}) and (\ref{DASC}), respectively.} 
\end{figure}

For the purposes of this work, let us consider three DTQWs, the first one with the shift operator $S$, the second one with the shift operator $T(\alpha)$, and the third one with the shift operator $V(\beta)$. In this work, we fix $\alpha=3\pi/16$ and $\beta=3\pi/8$. Later in this work, we will briefly refer to them as $S$, $T$, and $V$.

The walker position probability distribution $p(x, t)$ for each of the above DTQWs, understood respectively as DTQWs with the period-one sequence of the shift operators $S$, $T$, and $V$, has a bias toward negative values of $x$ [see Figs.~1(a), 1(b), and 1(c)] such that the walker average position $\ave{x}(t)$ is nonpositive [see Figs.~2(a), 2(b), and 2(c)].
Moreover, the walker position variance $\sigma^{2}(t)$ for each of the above DTQWs is ballistic, that is, it grows quadratically with time, $\sigma^{2}(t) \propto t^{2}$ [see Figs.~4(a), 4(b), and 4(c)].

\begin{figure}[t]
\centering
\begin{tabular}{cc}
\includegraphics[width=0.47\textwidth]{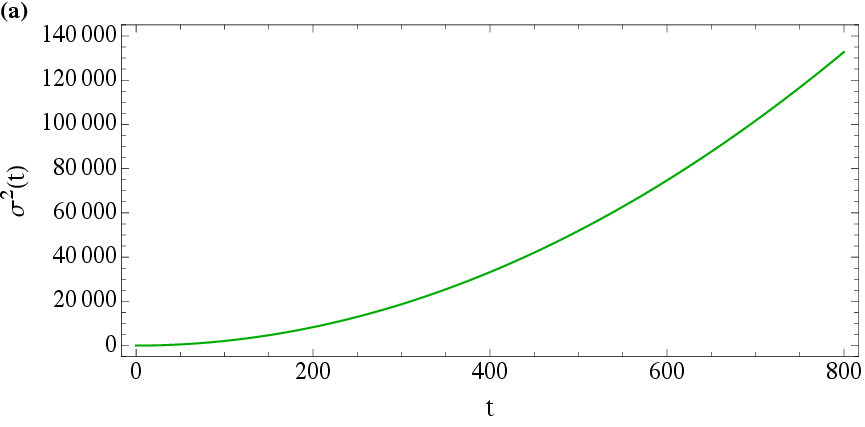} & \hspace{5mm} \includegraphics[width=0.47\textwidth]{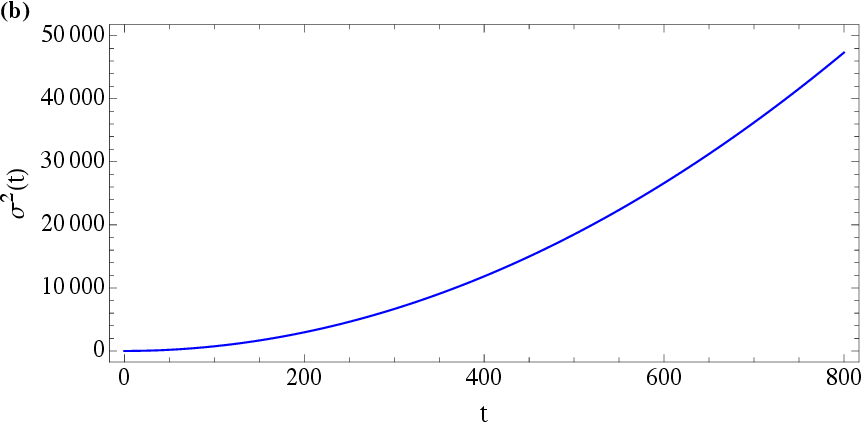} \vspace{5mm} \\
\includegraphics[width=0.47\textwidth]{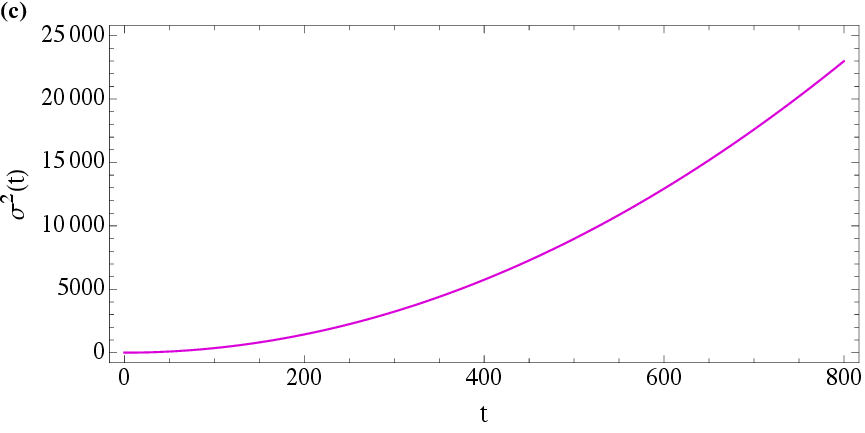} & \hspace{5mm} \includegraphics[width=0.47\textwidth]{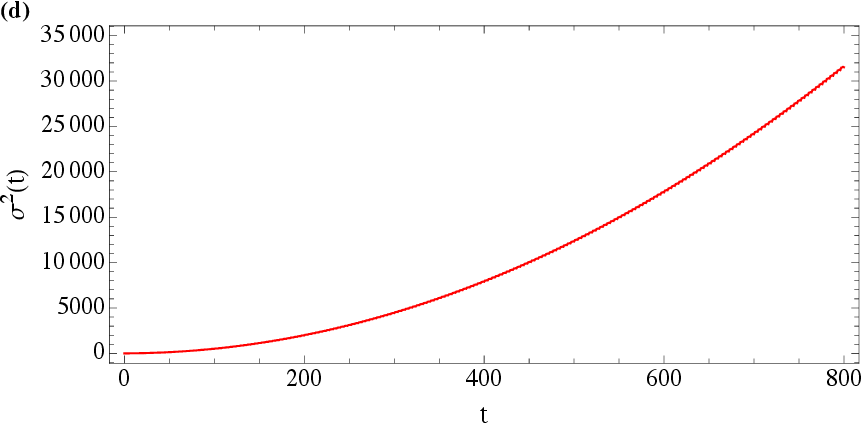} \vspace{5mm} \\
\includegraphics[width=0.47\textwidth]{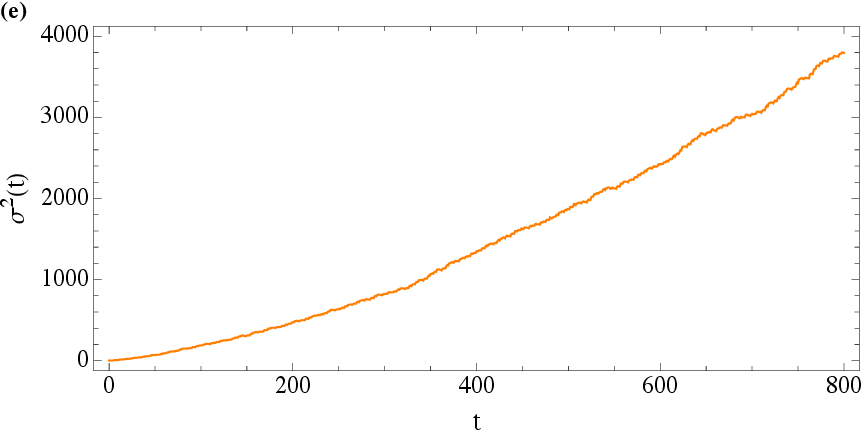}
\end{tabular}
\caption{The time evolution of the walker position variance. The plots (a), (b), and (c) are for 
DTQWs with the shift operators $S$, $T$, and $V$, respectively, understood as DTQWs with the period-one sequence of the shift operators $S$, $T$, and $V$, respectively. The plots (d) and (e) are for DTQWs with deterministic periodic and aperiodic sequences of the shift operators $S$, $T$, and $V$, given by Eqs.~(\ref{DPSC}) and (\ref{DASC}), respectively.} 
\end{figure}

\section{Parrondo's paradox for sequences of distinct shift operators}
Let us introduce now two DTQWs with the following deterministic sequences of the shift operators $S$, $T$, and $V$:
\begin{align}
\label{DPSC}
& SVTV~SVTV~SVTV~SVTV~SVTV  \dots,\\
\label{DASC}
& SVT~VSST~VSST~SVT~VSST~SVT~VSST \dots,
\end{align}
where spaces above have been added for better visibility of specific sequences of the shift operators. The first sequence is periodic with period $4$ (the string $SVTV$ is repeated) while the second sequence is aperiodic. The second sequence is obtained by placing the string $SVT$ at positions of nonprime numbers and the string $VSST$ at positions of prime numbers.
The walker position probability distribution $p(x, t)$ for each of the above DTQWs has a bias toward positive values of $x$ [see Figs.~1(d) and 1(e)] such that the average position of the walker $\ave{x}(t)$ is non-negative [see Figs.~2(d) and 2(e)]. 
Therefore, each of the above DTQWs displays Parrondo's effect.
Given the strength of this effect, it is similar to those presented in Refs.~\cite{Pires20,Walczak21,Walczak22} for Parrondo's effect in DTQWs with coin sequences. Moreover, the strength of Parrondo's effect for sequences of three distinct shifts considered here is comparable with its counterpart for sequences of two distinct shifts ~\cite{Walczak24}. 
The time evolution of the walker-coin quantum entanglement ${S}_{E}(t)$ for the above DTQWs [see Fig.~3(d) and 3(e)] shows that, in these cases, the walker-coin quantum entanglement is greater, on average, than in the cases of DTQWs with the period-one sequence of the shift operators $S$, $T$, and $V$, respectively [see Figs.~3(a), 3(b), and 3(c)]. 
Interestingly, the walker-coin quantum entanglement for DTQWs with coin sequences presented in Refs.~\cite{Pires20,Walczak21,Walczak22} exhibits a similar behavior. What is more, the walker-coin quantum entanglement for sequences of three distinct shifts is comparable with the case of sequences of two distinct shifts ~\cite{Walczak24}. Moreover, the walker position variance $\sigma^{2}(t)$ for each of the above DTQWs is ballistic [see Fig.~4(d) and 4(e)].

Given the substantial differences among the sequences of shift operators $S$, $T$, and $V$, as defined in Eqs.~(\ref{DPSC}) and (\ref{DASC}), the occurrence of Parrondo's effect in DTQWs with the initial state from Eq.~(\ref{IS}) is not immediately intuitive. There is no simple connection that clarifies this observation.

\section{Conclusion}
In this work, we have demonstrated the occurrence of Parrondo's effect in DTQWs employing deterministic aperiodic and periodic sequences of three distinct shift operators. These are the first examples of such DTQWs exhibiting this phenomenon. Notably, we have observed that Parrondo's effect, in this context, mirrors its influence on walker-coin quantum entanglement seen in DTQWs with varying coin sequences and a single standard shift operator. Specifically, Parrondo's effect enhances the walker-coin quantum entanglement. This is evidenced by comparing entanglement entropy approaching unity at large times, as shown in Fig.~3 of our prior works \cite{Walczak21,Walczak22} (for coin sequences), Fig.~3 of our previous work \cite{Walczak24} (for shift sequences), and Fig.~3 of this work (for shift sequences).

Furthermore, we have compared the magnitude of Parrondo's effect between these two scenarios. We assess this magnitude by examining the ratio of the absolute mean position, $|\ave{x}(t)|$, at large times $t$ for DTQWs with alternating sequences versus those with uniform sequences. By comparing Fig.~2 from our prior works \cite{Walczak21,Walczak22} (for coin sequences) with Fig.~2 from this work and our previous work \cite{Walczak24} (for shift sequences), we conclude that the strength of Parrondo's effect is at least comparable, if not greater, when using shift sequences compared to coin sequences.

Another thing worth emphasizing are non-standard shift (or ``translation'' \cite{Mastandrea24}) operators employed in the present work. In research on quantum walks, important issue is the control of transport and localization of the walker \cite{Ximenes26,Ximenes26b}. These properties depend on the initial state of the walker-coin system, coin operator and shift operator \cite{Romanelli09,deValcarel10,Panahiyan18}. For example, disorder in coined quantum walks usually leads to greater localization \cite{Vieira13,Vieira14,Yao23}. Let us note that two shift operators employed in the present work (see Eqs. (\ref{T}) and (\ref{V})) allow us to control the width of the probability distribution independently of the coin operator and initial conditions. For example, in the limits $\alpha \to 0$ and $\beta \to \pi/2$ one obtains $T(\alpha) \to ~{I}_{w} \otimes I_{c}$ and $V(\beta) \to {I}_{w} \otimes (\ketbra{0}{1}-\ketbra{1}{0})$, respectively. Hence, probability distributions in these limits should be very narrow (if the initial state is narrow). On the other hand, if $\alpha \to \pi/2$ and $\beta \to 0$ then probability distributions should be the widest. Let us also note that Eqs. (\ref{S16})-(\ref{S26}) allow for more complicated, multiparameter shift operators, that may be useful for optimizing the width of probability distributions.  

We anticipate that these results can be realized experimentally in quantum-optical platforms and may contribute to the development of novel quantum algorithms, including advanced protocols for generating highly entangled quantum states.

\section*{Declarations}

\begin{itemize}
\item This research received no external funding.
\item The authors declare no conflicts of interest.
\item Data sharing is not applicable to this article as no data sets were generated or analyzed during the current study. 
\item The numerical data presented in this study are available on request from the corresponding author.
\item Both authors contributed to the concept of the paper and analytical calculations. Z.W. made numerical calculations and figures. J.H.B. wrote the manuscript. Both authors have read and agreed to the published version of the manuscript.
\end{itemize}

\end{document}